\input harvmac
\def\newdate{Revised~12/8/2003}

\def\a{\alpha}
\def\b{\beta} \def\g{\gamma} \def\l{\lambda} \def\d{\delta} \def\e{\epsilon} \def\t{\theta} \def\r{\rho}

\def\s{\sigma} \def\O{\Omega}  \def\G{\Gamma}
\def\L{\Lambda} \def\D{\Delta}

\def\m{\mu}
\def\n{\nu}



 \lref\defsuperspace{
S.~Terashima and J.~T.~Yee,
{\it Comments on noncommutative superspace,}
arXiv:hep-th/0306237;
M.~Hatsuda, S.~Iso and H.~Umetsu,
{\it Noncommutative superspace, supermatrix and lowest Landau level,}
arXiv:hep-th/0306251;
S.~Ferrara, M.~A.~Lledo and O.~Macia,
{\it Supersymmetry in noncommutative superspaces,}
JHEP {\bf 0309}, 068 (2003)
[arXiv:hep-th/0307039];
T.~Araki, K.~Ito and A.~Ohtsuka,
{\it Supersymmetric gauge theories on noncommutative superspace,}
arXiv:hep-th/0307076;
R.~Britto, B.~Feng and S.~J.~Rey,
{\it Non(anti)commutative superspace, UV/IR mixing and open Wilson lines,}
JHEP {\bf 0308}, 001 (2003)
[arXiv:hep-th/0307091];
M.~Chaichian and A.~Kobakhidze,
{\it Deformed N = 1 supersymmetry,}
arXiv:hep-th/0307243;
A.~Sako and T.~Suzuki,
{\it Ring structure of SUSY * product and 1/2 SUSY Wess-Zumino model,}
arXiv:hep-th/0309076;
D.~Mikulovic,
{\it Seiberg-Witten Map for Superfields on Canonically Deformed N=1, d=4 Superspace,}
arXiv:hep-th/0310065.
}

\lref\renohalf{
R.~Britto, B.~Feng and S.~J.~Rey,
{\it Deformed superspace, N = 1/2 supersymmetry and (non)renormalization  theorems,}
JHEP {\bf 0307}, 067 (2003)
[arXiv:hep-th/0306215];
M.~T.~Grisaru, S.~Penati and A.~Romagnoni,
{\it Two-loop renormalization for nonanticommutative N = 1/2
supersymmetric WZ model,}
JHEP {\bf 0308}, 003 (2003)
[arXiv:hep-th/0307099];
A.~Romagnoni,
{\it Renormalizability of N = 1/2 Wess-Zumino model in superspace,}
arXiv:hep-th/0307209}
\lref\REY{
R.~Britto and B.~Feng,
{\it N = 1/2 Wess-Zumino model is renormalizable,}
arXiv:hep-th/0307165;
O.~Lunin and S.~J.~Rey,
{\it Renormalizability of non(anti)commutative gauge theories with N = 1/2 supersymmetry,}
JHEP {\bf 0309}, 045 (2003)
[arXiv:hep-th/0307275];
D.~Berenstein and S.~J.~Rey,
{\it Wilsonian proof for renormalizability of N = 1/2 supersymmetric field theories,}
arXiv:hep-th/0308049.
}

\lref\AlishahihaKG{
M.~Alishahiha, A.~Ghodsi and N.~Sadooghi,
arXiv:hep-th/0309037.
}


\Title{\vbox{\baselineskip12pt
\hbox{YITP-SB-03-55}
}}
{\vbox{
\centerline{Instanton Calculations for}
\vskip  .3cm
\centerline{$N={1\over 2}$ super Yang-Mills Theory}
}}
\smallskip
\medskip\centerline{P.A.
Grassi$^{~a,b,}$\foot{pgrassi@insti.physics.sunysb.edu},
R. Ricci$^{~a,}$\foot{rricci@grad.physics.sunysb.edu}, and D.
Robles-Llana$^{~a,}$\foot{daniel@insti.physics.sunysb.edu}}
\medskip
\centerline{$^{(a)}$ {\it C.N. Yang Institute for Theoretical Physics,}
}
\centerline{\it State University of New York at Stony Brook,
NY 11794-3840, USA}
\medskip
\centerline{$^{(b)}$ {\it Dipartimento di Scienze,
Universit\`a del Piemonte Orientale,}}
\centerline{\it
C.so Borsalino 54, Alessandria,  15100, ITALY}

\bigskip
\vskip 1cm
\noindent

We study (anti-) instantons in super Yang-Mills theories defined
on a non anticommutative superspace. The instanton solution that
we consider is the same as in ordinary $SU(2)$ $N=1$ super
Yang-Mills, but the anti-instanton receives corrections to the
$U(1)$ part of the connection which depend quadratically on
fermionic coordinates, and linearly on the deformation parameter
$C$. By substituting the exact solution into the classical
Lagrangian the topological charge density receives a new
contribution which is quadratic in $C$ and quartic in the
fermionic zero-modes. The topological charge turns out to be
zero. We perform an expansion around the exact classical solution
in presence of a fermionic background and calculate the full
superdeterminant  contributing to  the one-loop partition function. We find that the one-loop partition function is not modified with respect to the usual $N=1$ super Yang-Mills.
 \Date{\newdate}


\baselineskip16pt

\newsec{Introduction}

It is a common belief that in the standard formulation of
superspace, only the bosonic subspace may have a non-trivial
topology. A superspace ${\cal M}^{(n|m)}$, where $n$ is the
dimension of its bosonic subspace ${\cal M}_{b}$ and $m$ is the
dimension  of spinor representation, is a Grassmanian vector
bundle with no topology in the fibers. However, despite some
attempts to construct models with non trivial superspace topology
(see for example \lref\SchwarzPF{ J.~H.~Schwarz and P.~Van
Nieuwenhuizen, {\it Speculations Concerning A Fermionic
Substructure Of Space-Time,} Lett.\ Nuovo Cim.\  {\bf 34}, 21
(1982). } \SchwarzPF, \lref\BouwknegtMN{ P.~Bouwknegt,
J.~G.~McCarthy and P.~van Nieuwenhuizen, {\it Fusing the
coordinates of quantum superspace,} Phys.\ Lett.\ B {\bf 394}, 82
(1997) [arXiv:hep-th/9611067]. } \BouwknegtMN\ and \lref\BarsTU{
I.~Bars and S.~W.~MacDowell, {\it Spinor Theory Of General
Relativity Without Elementary Gravitons,} Phys.\ Lett.\ B {\bf
71}, 111 (1977). } \BarsTU) and interesting arguments suggesting that only the topology of the bosonic subspace really
matters (see for example \lref\gates{ S.~J.~Gates, {\it
Ectoplasm has no topology: The prelude,} arXiv:hep-th/9709104\semi
S.~J.~Gates, {\it Ectoplasm has no topology,} Nucl.\ Phys.\ B
{\bf 541}, 615 (1999) [arXiv:hep-th/9809056].} \gates), a new
superspace formulation
 \lref\sei{
N.~Seiberg, {\it Noncommutative superspace, N = 1/2
supersymmetry, field theory and  string theory,} JHEP {\bf 0306},
010 (2003) [arXiv:hep-th/0305248]. N.~Berkovits and N.~Seiberg,
{\it Superstrings in graviphoton background and N = 1/2 + 3/2
supersymmetry,} JHEP {\bf 0307}, 010 (2003)
[arXiv:hep-th/0306226]. } \sei\ based on a construction in superstring theory
\lref\noncom{ J.~de Boer, P.~A.~Grassi and P.~van Nieuwenhuizen,
{\it Non-commutative superspace from string theory,} Phys. Lett.
B. [arXiv:hep-th/0302078]; H.~Ooguri and C.~Vafa, {\it The
C-deformation of gluino and non-planar diagrams,}
[arXiv:hep-th/0302109]; H.~Ooguri and C.~Vafa, {\it Gravity
induced C-deformation,} arXiv:hep-th/0303063. } \noncom\foot{A
recent work \lref\BarsDQ{ I.~Bars, C.~Deliduman, A.~Pasqua and
B.~Zumino, {\it Superstar in noncommutative superspace via
covariant quantization of the superparticle,}
arXiv:hep-th/0308107. } \BarsDQ\ shows that a deformed superspace
structure also emerges in superparticle theory.}, reopened the
debate. For related considerations on deformed superspaces see also 
\lref\KlemmYU{
D.~Klemm, S.~Penati and L.~Tamassia,
Class.\ Quant.\ Grav.\  {\bf 20}, 2905 (2003)
[arXiv:hep-th/0104190].
} \KlemmYU, 
\lref\TerashimaRI{
S.~Terashima and J.~T.~Yee,
arXiv:hep-th/0306237.
} \TerashimaRI.

In this approach, the fermionic coordinates $\t^{\a},
\bar\t^{\dot \a}$ are no longer Grassmann variables, but they are
promoted to elements of a Clifford algebra \eqn\ooA{ \{ \t^{\a},
\t^{\b} \} = 0\,, ~~~~~ \{ \t^{\a}, \bar\t^{\dot\b} \} = 0\,,
~~~~~ \{ \bar\t^{\dot\a}, \bar\t^{\dot\b} \} = C^{\dot\a \dot\b}
\,. } where $C^{\dot\a \dot\b}$ is the constant self-dual RR field strengh
of the closed string theory background. As a consequence the $N=1$ supersymmetry algebra is deformed and broken down to $N=1/2$ \sei.

From a more physical perspective, and after several perturbative studies of $N=1/2$ supersymmetric quantum field theories \renohalf, one is tempted to ask about their non-perturbative aspects. The issue is not unrelated to the problem addressed in the previous paragraph: it is by now well established that the main sources of non-perturbative physics are objects which have also a special topological significance. One would then hope that knowing more about the non-perturbative regime of $N=1/2$ supersymmetric theories might in addition shed some light on a possible non-trivial topology of superspace. In particular, in this paper we study instantons (anti-instantons), {\it i.e.} finite-action anti-selfdual (self-dual) solutions to the Euclidean equations of motion of (super) Yang-Mills theories, which have proven to be one of the main sources of insights in both the non-perturbative regime of quantum field theories, and the topology of four-manifolds (for a physics review, see for example \lref\ShifmanMV{
M.~A.~Shifman and A.~I.~Vainshtein, {\it Instantons versus
supersymmetry: Fifteen years later,} arXiv:hep-th/9902018. }
\ShifmanMV; for a mathematical introduction see \lref\FreedXE{
D.~S.~Freed and K.~K.~Uhlenbeck,
``Instantons And Four - Manifolds,'' Second Edition, Springer Verlag, 1991.} \FreedXE). As is well-known, the instanton
charge is topological and completely computable in terms of the bosonic solution to
the self-dual Yang-Mills equations. Moreover, instantons are
degenerate solutions, and it is crucial to study their moduli space,
 which  is parametrized by a set of variables which are referred to
 as collective coordinates. In some special instances a complete
  parametrization of this space can be obtained through the
  generators of the symmetries of the equations of motion
  which are broken by the classical solution (an example is $N=1$ SYM with $SU(2)$ gauge group).
   More generally, however, one has to find the most general solution to the
    equations of motion through the ADHM construction
    (see for example \lref\instquatt{ A.~V.~Belitsky,
S.~Vandoren and P.~van Nieuwenhuizen, {\it Yang-Mills and
D-instantons,} Class.\ Quant.\ Grav.\  {\bf 17}, 3521 (2000)
[arXiv:hep-th/0004186]} \lref\instquattB{N.~Dorey, T.~J.~Hollowood, V.~V.~Khoze
and M.~P.~Mattis, {\it The calculus of many instantons,} Phys.\
Rept.\  {\bf 371}, 231 (2002) [arXiv:hep-th/0206063]. }
\instquatt, \instquattB\ and references therein).

Supersymmetry adds many interesting features to the study of instantons. The most salient is perhaps the fact that instantons in supersymmetric theories break half of the supersymmetries of the original action, in addition to translations, dilatations, and half of the Lorentz symmetry (this is possible only in Euclidean space). To be more specific, instantons
($F^{+}_{\mu\nu} =0$) break the supersymmetries generated by $\bar
Q^{\dot\a},S^{\a}$, while anti-instantons ($F^{-}_{\mu\nu} =0$) break $Q^{\a},\bar S^{\dot\a}$ \foot{We are using the notations of \ShifmanMV\ for the generators of the super-conformal group.}. These broken supercharges give rise to fermionic collective coordinates, which can be thought of as the fermionic superpartners of the bosonic coordinates introduced above. Again, finding the complete set of fermionic collective coordinates requires solving the full equations of motion. In geometrical terms, the fermionic collective coordinates can be seen to parametrize the symplectic tangent space to the moduli space.

When considered from the quantum field theoretical perspective, instantons characterize topological vacua of the Euclidean theory around which one must expand in the computation of the path integral. Going back to Minkowski space they give the main contribution to tunneling processes which go as the square of the inverse of the coupling constant, and thus can never be seen in ordinary perturbation theory. In the semi-classical approximation, one must expand the classical action up to quadratic terms in quantum fluctuations around the instanton. The measure in the path integral is then modified by the degeneracy of solutions, which translates into the presence of zero-modes in the functional determinants. The correct normalization is determined  by the jacobian obtained by trading the integration over the zero-modes for the collective coordiantes. As we saw, in super Yang-Mills theories, instantons have fermionic counterparts which depend on fermionic collective coordinates. These must be accounted for by including the Pfaffian associated to their inner products in the measure, and the Grassmann variables are to be integrated over using Berezin integration. This leads to new genuinely non-perturbative effects, such as the well-known non-vanishing result for the gluino condensate.

In this paper we would like to see how the characteristics of instanton calculations in ordinary supersymmetric theories are modified when one considers quantum field theories defined in deformed superspace. We will concentrate on pure $U(2)$ $N=1/2$ SYM, as an already non-trivial illustration of how these differences arise.

First and foremost, in $N=1/2$ theories the commutation relations for half of the supercharges, say $Q^{\a}$,
are deformed to \eqn\defA{ \{Q_{\a}, Q_{\b} \} = - 4
C^{\dot\a\dot\b} \s^{\m}_{\a\dot \a} \s^{\n}_{\b\dot\b}
{\partial^{2} \over \partial y^{\m} \partial y^{\n}}\,. }
The Lagrangian for pure $N=1/2$ SYM contains two additional operators of dimensions 5 and 6 with respect to the $N=1$ case \sei, but  it is nonetheless renormalizable \renohalf\ (see later).

The fields obeying the anti-self-duality equations of conventional SYM ($F^{+}_{\m\n}=0$) furnish a complete solution to the equations of motion even in the presence of the extra couplings. Moreover, the supersymmetries broken by these solutions are $\bar Q_{\dot \a}$,$S_{\a}$, which are not broken by the C-deformation. Consequently, the subalgebra generated by these supercharges should lead to the complete set of collective coordinates
as usual, and the path integral measure can be constructed in the
conventional way, at least at the classical level. The classical instanton action is the same as in $N=1$ SYM. The fermionic zero-modes of the Dirac operator in the instanton background could however be lifted in perturbation theory due to the presence of extra non-supersymmetric couplings in the action, thereby giving corrections to the effective action at one-loop.

On the other hand, $N=1$ anti-instantons
($F^{-}_{\mu\nu}=0$) do not provide a full solution to the
classical equations of motion of the deformed theory. As noticed in \lref\imA{A.~Imaanpur, {\it On instantons and zero modes of N = 1/2 SYM theory,} JHEP {\bf
0309}, 077 (2003) [arXiv:hep-th/0308171]. } \imA\  the extra couplings contribute fermionic source terms for the Dirac equation in the anti-instanton background. In this paper we revisit the procedure of \imA\  and solve the equations of motion exactly. Our perspective, however, differs from that of \imA, and is more in line with the overall philosophy of \instquatt. We derive the complete solution through an iterative procedure which consists in systematically expanding the equations of motion in powers of the fermionic quasi-collective coordinates. In the end we arrive at the conclusion that the ordinary $SU(2)$ supersymmetric anti-instanton is supplemented with a non-trivial $U(1)$ connection which depends quadratically on fermionic variables (the Grassmann collective coordinates of the $N=1$ solution). When we substitute this solution into the lagrangian we find the density charge
\eqn\Szero{
\Delta {\cal L}_{anti-instanton}= {3072\r^6 (2x^4-3\r^2 x^2)\over(x^2 +\r^2)^6}|C|^2 \bar\eta^2\xi^2}

The new feature of \Szero\ is the fact that it depends on the
fermionic parameters $\xi_{\a}$ and $\bar\eta^{\dot\a}$. The appearance of the second term in
the exponent of \Szero\ can be traced back to the fact that the
anti-instanton breaks the supersymmetries which are already
broken by the C-deformation. Consequently the fermionic
parameters entering the anti-selfdual solution cannot be viewed
as collective coordinates in the usual sense. Notice however that
performing the integration over the bosonic variables $x$, the
total charge, namely the instanton action, reproduce the usual
undeformed $N=1$ answer (in the text, an argument based on the
Atiyah Singer index Theorem will be used) \foot{This is consistent with the considerations of \gates.}.

A similar situation occurs in $N=4$ SYM. In that case, not only
the charge density, but also the total charge depends upon the
fermionic ``quasi-collective coordinates'', but this relies on
the approximate nature of the solutions of the equations of
motion \instquatt, whereas in the present case the appearance of
zero-modes in the Lagrangian is a direct consequence of the broken
supersymmetry of the classical action \foot{A quantum mechanical
 situation in which exact solutions to the equations of motion
 yield an action depending on collective coordinates can be found
 in \lref\van{S.~Vandoren and P.~van
Nieuwenhuizen, {\it New instantons in the double-well potential,}
Phys.\ Lett.\ B {\bf 499}, 280 (2001) [arXiv:hep-th/0010130].}
\van.}. Note that, as the C-deformation also breaks dilatational symmetry, it is perfectly reasonable that the parameter $\r$, which labels the size of the anti-instanton,  appears in \Szero. 

The second step in path integral computations around a classical
solution is the evaluation of radiative corrections. In the
conventional case, the unbroken supersymmetries pair up the
bosonic and fermionic fluctuations in the one-loop determinant,
with the net effect of only a prefactor in the path integral
depending on the regulator mass $\m$. Together with the measure
coming from the zero modes, it encodes the one-loop
renormalization of the coupling constant. In section 6 we will
discuss the one loop corrections to the effective action for
N=1/2 SYM in anti-instanton background. The novelty is the
treatment of the fermionic modes. As discussed in \instquatt\  one
can either treat the fermions perturbatively, expanding around a
purely bosonic configuration, or {\it exactly} including them
already in the classical background. Here we follow the second
route. This has the following important consequence for the
one-loop
 calculation: Performing an expansion around the full background of the C-deformed action
 induces new bilinear couplings between bosonic and fermionic
 fluctuations. One is then prompted to calculate a superdeterminant
 in the space of all fields (unlike the conventional situation in
  which bosonic and fermionic determinants decouple).
   This computation, to the best of our knowledge, has never
    before been performed in this context (see however 
\lref\JackVJ{
I.~Jack and H.~Osborn,
Nucl.\ Phys.\ B {\bf 249}, 472 (1985).
} \JackVJ). The one loop effective action turns out
to be zero also. The physical reason behind the cancellation has
to be found in the fact that the deformation in the
 sector considered breaks exactly the same supersymmetry
 generators broken already by the anti-instanton
 solution. Therefore we are left with the same supersymmetry
 generators as in N=1 with anti-instanton backgorund. The pairing mechanism which is responsible
 for the cancellation of the one loop contribution in the
 familiar N=1 case turns out to be effective also in the present situation.

From this result we can derive interesting consequences. As usual in super-instanton computations one can obtain the the
renormalization group $\b$-function by combining the measure
for zero-modes with the one-loop contribution to the path-integral. We find that the part independent of the Grassmann variables in the effective action gives the usual running of the coupling constant for $U(2)$ super Yang-Mills. 

The gluino condensate is not deformed by $C$ since the
anti-instanton moduli space measure is unchanged with respect to
the N=1 case.

Finally one word of caution is in order as to the physical
interpretation of our results. This model makes sense only in
Euclidean spacetime and therefore the usual physical
interpretation of (anti)-instanton solutions as tunneling
processes between topologically different bosonic Minkowski vacua
of the theory is elusive.

The paper is organized as follows: in Sec. 2, we recall
 the basic facts about deformed superspace
and super Yang-Mills theory. In Sec. 3, we discuss the role of collective coordinates for instanton and
anti-instanton in deformed superspace. In Sec. 4, we solve the equations of motion iteratively
 for both instanton and (anti)-instanton. In Sec. 5, we compute the classical Lagrangian. In Sec. 6 we derive
  the (quasi) zero-mode measure,
 and the one-loop contributions to the effective action.
 Conclusions and future directions are given in Sec. 7.

\newsec{The action and the symmetries}

The lagrangian in deformed superspace for general $U(N)$ gauge
group is \eqn\ACT{ {\cal L} = - {\rm tr} \Big[{1\over 2}F_{\m\n}
F^{\m\n} + 2i \l \not\!\!D \bar\l-  D^{2} - i g
C^{\m\n} F_{\m\n} \l \l  - g^{2} {|C|^{2}\over 4} (\l \l)^{2} \Big]
\,.} Note that our conventions are opposite to the ones in \sei. We have chosen anti-hermitian generators {\it i.e.} for $U(2)$
we take $T^{a}=i{\s^{a}\over 2}$ for the $SU(2)$ subgroup and
$T^{4}={i\over 2}$ for the $U(1)$ part. In this way ${\rm tr}
\{T^{a}T^{b}\}=-{1\over 2} \d^{ab}$. $C^{\m\n}=C_{\dot\a \dot\b}
\e^{\dot\b \dot\g} {{\bar\s}^{\m\n \dot\a}}_{\dot\g}$ is
anti-symmetric and anti-selfdual. In these conventions the
covariant derivative for any field in the adjoint is
$D_{\m}=\del_{\m} + g [A_{\m},\cdot]$.

For a generic group, we have that $\bar\l_{\a} \bar\l_{\b}
\e^{\a\b} = {1\over 2}\bar\l^{a}_{\a} \bar\l^{b}_{\b} \e^{\a\b}
\{T^{a}, T^{b} \}$ and in the case of $G= SU(N)$, this is equal
to ${1\over 2}\bar\l^{a}_{\a} \bar\l^{b}_{\b} d^{a b c} \e^{\a\b}
T^{c}$ \foot{It should be noted that, in contrast to $N=1$ SYM,
the tensor $d^{a b c}$ enters the classical action. The same
phenomenon is present in non-commutative space-time (see
\lref\SeibergVS{ N.~Seiberg and E.~Witten,
JHEP {\bf 9909}, 032 (1999)
[arXiv:hep-th/9908142].
} \SeibergVS\ and references thereof).}. This is clearly zero for $SU(2)$.  The action is not
hermitian and it does not preserve $R$ symmetry. Notice that the
operators in the second line on \ACT\ have dimensions 5 and 6.
Nevertheless it turns out that the action is renormalizable
\renohalf. The two operators break conformal invariance of the
classical theory since the constant $C^{\a\b}$ has mass dimension
-1. It has been shown in \REY\ that unusual mass dimensions can be assigned to $\l$, $\bar\l$ and $C$. This is the key to the renormalizability by power counting.

We work in chiral superspace with $y^{\m}=x^{\m}-i\t\s^{\m}\bar{\t}$
and supercharges
\eqn \rc{Q_{\a}={\del\over
\del{\t}^{\a}}-2i\s^{\m}_{\a\dot{\a}}\bar{\theta}^{\dot{\a}}{\del\over
\del y^{\m}} \,, ~~~~~~~
\bar{Q}_{\dot{\a}}=-{\del\over \del\bar{\t}^{\dot\a}}.}
In the supersymmetry algebra only the
anticommutator of the $Q_{\a}$'s gets modified when we turn on $C^{\dot\a \dot\b}$
\eqn\rb{\{Q_{\a},Q_{\b}\}=-4C^{\dot{\a}\dot{\b}}\s_{\a\dot{\a}}^{\m}\s_{\b\dot{\b}}^{\n}
{{\del}^2\over \del y^{\m}\del y^{\n}}.} The explicit presence of
the $C$ deformation in the algebra breaks the amount of
supersymmetry from N=1 to N=1/2. The only preserved charges are
the $\bar{Q}^{\dot{\a}}$'s.

 The symmetry of the action for $C=0$
is the complete superconformal group generated by $D, \Pi,
Q_{\a}, \bar Q_{\dot\a}, S_{\a}, \bar S_{\dot \a}, P_{\a\dot \a},
M_{\a\b}, \bar M_{\dot \a\dot\b}$ where $\Pi$ is the generator
of  $U(1)$ R-symmetry and $D$ is the generator of the dilatations.
Turning on the parameter $C^{\dot\a \dot\b}$, the
symmetry group is broken to the group generated by $Q_{\a},
P_{\a\dot\a}, M_{\dot \a\dot \b}, \bar S_{\dot \a}$ which form a
subalgebra.


\newsec{Collective Coordinates}

We now describe the collective coordinates which parametrize the
coset space $G/H$ where $G$ and $H$ are the symmetry groups of
the action and of the solution, respectively. Therefore $G/H$
represents the symmetries of the action which are broken by an
explicit instanton solution. In $N=1$ $SU(2)$ super Yang-Mills, given a bosonic solution to the
equations of motion one can reconstruct the most general solution
depending on the set of bosonic and fermionic collective
coordinates $\{b_{i}\}, \{ f_{i}\}$ applying the generalized
shift operator $V(\{b_{i}\},\{f_{i}\})= \prod_{i,j}
e^{Q^{bos}_{i}b_{i}} e^{Q^{ferm}_{j}f_{j}}$ where $Q^{bos}_{i}$
and $Q^{ferm}_{j}$ are the fermionic and bosonic generators of
broken symmetries \ShifmanMV. Let  us begin considering the
deformed anti-instanton solution \imA.  The usual solution is
modified by the presence of Grassmann variables. The novelty here is
that the bosonic background breaks the chiral supersymmetries
which are already broken by the C deformation from the outset.  The axial R symmetry generated by
  $\Pi=\t^{\a}\del/\del{\t}^{\a}-\bar{\t}^{\dot{\a}}\del/\del\bar{{\t}}^{\dot{\a}}$
  is also explicitly broken by the C deformation
  $\{\t^{\dot{\a}},\t^{\dot{\b}}\}=C^{\dot{\a}\dot{\b}}$. The R symmetry
  can be restored introducing the term $-2C^{\a\b}\del/\del
  C^{\a\b}$ in $\Pi$ assigning R symmetry number -2 to
  $C^{\a\b}$ \foot{This is the usual procedure of treating the parameters in the classical action as constant background fields, and assigning spurious charges to them \lref\IntriligatorAU{
K.~A.~Intriligator and N.~Seiberg,
Nucl.\ Phys.\ Proc.\ Suppl.\  {\bf 45BC}, 1 (1996)
[arXiv:hep-th/9509066].} \IntriligatorAU.}
As already remarked the dilatations are broken: the
  classical action is not conformally invariant. It is also easy to
  check that only the chiral superconformal generators $S_{\a}$'s are symmetries.
  The Lorentz invariance of the action is also broken down to the
  Lorentz generators $M_{\a\b}$. The symmetry group of the Lagrangian is therefore
  $G=(\bar{Q}_{\dot{\a}},S_{\a},P_{\m},M_{\a\b})$.

 The symmetry group preserved by the anti-instanton solution is
 $H=(\bar{Q}_{\dot{\a}},S_{\a},M_{\a\b})$ and
  $G/H=\{P_{\m}\}$.  The moduli space of the instanton solution is
  therefore parameterized by the collective coordinate corresponding to
  the translations. In
  particular the Grassmann parameters $\xi_{\a}$ and
  $\bar{\eta}_{\dot{\a}}$ which, in the usual $N=1$ superspace,
  parameterize the moduli space do no longer index exact
  zero modes of the action. We therefore expect them to appear in the classical action once we substitute the exact solution. Still they must be integrated over in the path integral, with the modified measure given by their potential in the classical action.

In the instanton background the group of symmetries preserved by the solution is
   $H=(Q_{\a},\bar{S}_{\dot{\a}},\bar{M}_{\dot{\a}\dot{\b}})$.
   Therefore we have the usual coset
   $G/H=(\bar{Q}_{\dot{\a}},S_{\a},P_{\mu},M_{\a\b})$.

\newsec{Iterative solution of the equations of motion}

In this section we solve the classical equations of motion (we restrict our attention to $U(2)$ gauge group) order by order in fermionic quasi-collective coordinates. We will see that the usual anti-instanton ($F_{\m\n}^{-}=0$) receives corrections of order $g^{0}$ quadratic in fermionic coordinates, whereas the instanton ($F_{\m\n}^{+}=0$) remains an exact solution. The equations of motion for the lagrangian \ACT\ are
\eqn\B{
D^{\mu} (F_{\mu\nu} - i g C_{\mu\nu} \l \l) + i (\bar\sigma_{\a \dot\a})_\nu \{\l^{\a},\bar\l^{\dot\a} \} = 0\,,
}
$$
i \s^{\mu} D_{\mu} \bar\l =   \l \Big(i g C^{\m\n} F_{\mu\nu}^{-} +g^{2} {{|C|^{2}}\over 2}  \l \l \Big)\,,~~~~~
 \bar\s^{\mu} D_{\mu} \l = 0\,.
$$

\subsec{Anti-instanton solution}

We want to solve \B\ in an anti-instantonic background. Anticipating the final result, and as a kind of synopsis, we can already say that the fields admit an expansion \eqn\nnn{\eqalign{ &
A_{\m}=g^{-1}A_{\m}^{(0)}+g^{0}A_{\m}^{(1)} \cr & \l = g^{-1}
\l^{(0)},~~~~~~~ \bar\l=0 }\,.}

We start by setting the fermions to zero,
$\l = \bar\l =0$. The equations become \eqn\C{
D^{\m}F_{\m\n} = 0\,.} As customary, this equation admits self-dual
solutions for which \eqn\D{ F_{\m\n}^{(0),-} = 0\,.} The gauge field
configuration which solves \D\ is the usual anti-instanton
\eqn\antiinstanton{\eqalign{& A^{a}_{\mu}(x,x_{0},\r) = {1\over g} { {2
{\eta}_{\m\n}^{a}(x-x_{0})_{\n}} \over (x-x_{0})^{2} + \r^{2}} =
g^{-1} A_{\m}^{(0),a}\cr
& F_{\m\n}^{(0),a}(x,x_{0},\r)={1\over g} \eta_{\m\n}^{a}{\r^{4}\over [(x-x_0)^2 + \r^2]^4}}} where the index $a$  belongs to $SU(2)$,
and ${\eta}_{\m\n}^{a}$ are the 't Hooft symbols. We have written \antiinstanton\ in regular gauge. Note that at
this stage the $U(1)$ part of the connection is zero, as usual
$U(1)$ (anti-) instantons are necessarily flat. However, we must
substitute the background \antiinstanton\ into the Dirac
operator. The equations for the fermions become \eqn\fermions{
\sigma^{\m} D_{\m}^{(0)} \bar\l^{(0)}= 0,~~~~~~~~~ \bar{\sigma}^{\m}
D_{\m}^{(0)} \l^{(0)} = 0 } where $D_{\m}^{(0)}$ is the Dirac operator
with respect to the connection \antiinstanton. Now, the second
equation in \fermions\ has non-trivial zero-modes given by
\eqn\E{ \l^{\a} = - {1\over 2} {{\s_{\m\n}}^{\a}}_{\b}
\Big( \xi^{\b} - \bar\eta_{\dot\gamma} {{\bar\s_{\r}^{\dot\gamma
\b}}}(x-x_{0})^{\r} \Big) F_{\m\n}^{(0)} = g^{-1} \l^{(0),\a}}
where $F_{\m\n}^{(0)}= F_{\m\n}^{(0),a}T^a$. As an aside, note that the modes \E\ are not
generated by any symmetry of the equations of motion, as
superconformal transformations are explicitely broken in this
sector. Nevertheless their presence is required by the index
theorem.

Once $\l^{(0)}$ has been switched on the equations of motion are
modified to \eqn\F{\eqalign{ & 0 = D^{(0),\m}(F_{\m\n}^{(1)} -  i
{C_{\m\n}\over g} \l^{(0)} \l^{(0)}) \cr & 0 =  i \s^{\mu} D^{(0)}_{\mu}
\bar\l  - i \l^{(0)} C_{\m\n} \Big( {F^{(1)}_{\mu\nu}} -i
{{C_{\m\n}}\over 2g} \l^{(0)} \l^{(0)} \Big) \,.}} In order to
satisfy both equations simultaneously, notice that $C_{\m\n}$ is
anti-selfdual, and thus $C_{\m\n}F_{\m\n}=C_{\m\n}F_{\m\n}^{-}$.
Using this and the Bianchi identity
$D^{\m}F_{\m\n}^{+}=D^{\m}F_{\m\n}^{-}$ we see that one can set
$\bar\l=0$ and impose \eqn\G{ F^{{(1)},-}_{\m\n} - {i\over
2g} C_{\m\n} {\l^{(0)}} {\l^{(0)}} = 0\,. } The analysis of \imA\
shows that the correction to $F^{-}_{\m\n}$ to second order in
Grassmann coordinates affects only the $U(1)$ part of the
curvature, and takes the form \eqn\H{ F_{\m\n}^{(1), 4-} = -
{1\over 4g } C_{\m\n} {\l}^{(0), a} {\l}^{(0),a} } with $\l^{4}$=0.
Using some spinor algebra, we can calculate the square on the
right-hand side. We find \eqn\I{ {1\over 2}{\l}^{(0),\a
a}{\l}^{(0), a}_{\a} = 96 \Big[{\xi}^{2} - 2 \bar\eta \bar\s_\r
\xi (x-x_{0})^{\r} + {\bar\eta}^{2} (x-x_{0})^{2} \Big]
{\r^{4}\over [(x-x_{0})^{2}+{\r}^{2}]^{4}} \,.} We see that the
$U(1)$ part of the connection is of order $g^{0}$ and depends
quadratically on the Grassmann coordinates.

For later use we
write down explicit expressions for the different Grassmann
components of $A^{(2),4}_{\m}$. The strategy to follow is based
on the fact that if $F_{\m\n}^{-4}$ can be written as $BC_{\m\n}
\del^{2}K(x)$ then $A_{\m}^{4}=-BC_{\m\n}{\del}_{\n}K(x)$, where
$B$ is a constant. A simple way to prove this is using spinor
notation: assuming
$A_{\a\dot{\a}}=C_{\a}^{\g}\del_{\g\dot{\a}}K$. Then \eqn\ab
{F_{\a\b}=\del_{(\a}^{\dot{\a}}A_{\b)\dot{\a}}=C_{(\b}^{\g}\del_{\a)}^{\dot{\a}}
\del_{\g\dot{\a}}K\sim C_{\a\b}{\del}^2K} where we used
$\del_{\a}^{\dot{\a}}\del_{\g\dot{\a}}={1\over 2}
\epsilon_{\g\a}\del^2$.

 In our case we have \eqn\J{ F_{\m\n}^{(\xi)^{2},4-} = -2
{\xi^2} C_{\m\n} \del^{2} K_{1}(x,x_{0},\r) }
$$
F_{\m\n}^{({\bar\eta}^{2}), 4-} = 2 \r^{2} {\bar\eta}^{2} C_{\m\n} \del^{2} K_{2}(x,x_{0},\r)
$$
$$
F_{\m\n}^{(\bar\eta \xi), 4-} = 4 \r^{2} C_{\m\n} \del^{2} K_{3}(x,x_{0},\r)
$$
where
\eqn\K{\eqalign{
& K_{1}(x,x_{0},\r) = {(x-x_{0})^{2}\over [(x-x_{0})^{2}+\r^{2}]^{2}} - {2\over (x-x_{0})^{2}+\r^{2}} \cr
& K_{2}(x,x_{0},\r) = {(x-x_{0})^{2}\over [(x-x_{0})^2 + {\r^2}]^2} + {1\over (x-x_{0})^2 + {\r^2}} \cr
& K_3(x,x_{0},\r) = {{\bar\eta \bar\s_\r \xi}(x-x_{0})^{\r}\over [(x-x_{0})^2 + \r^2]^2}\,.
}}
It is easy to see that the iteration procedure stops at order ${C\over g}$: the covariant
 derivative in $SU(2)$ does not get corrections in Grassmann variables, and the
 covariant derivative in $U(1)$ is the normal derivative for fields in the adjoint,
 and is thus insensitive to the Grassmann-corrected part of the $U(1)$ connection.
 Therefore there are no further $U(1)$ normalizable zero modes and $\l^{4}$ stays
 zero at further orders in the coupling constant
  \foot{Thinking of the $C$ parameter as a smooth deformation we can
  still apply the Atiyah-Singer Theorem
  \eqn\rh{ind(\not\!\!D)=-{1\over {4{\pi}^2}}\int {\rm Tr}F\wedge F }
       Indeed notice that the index of the Dirac operator (which measures the difference between
 the number of zero-modes for $\l_{\a}$ and $\bar \l_{\a}$)
   does not depend on the deformation parameter C since the equation
   for ${\bar\l}$ is the usual
   Dirac equation and the equation for $\l$ becomes the usual
  one when we impose the modified instanton condition. From this we
infer that there are no zero modes associated to the $U(1)$ part
of the connection and the equality \rh forces the additional U(1)
piece (U(1)topological charge) $\int
  F_{\m\n}^4F_{\m\n}^4$ to vanish.}. The full solution can then be written
\eqn\fullantiinstanton{\eqalign{
& A_{\m} = g^{-1} { {2 {\eta}_{\m\n}^{a}T^{a}(x-x_{0})_{\n}} \over
(x-x_{0})^{2} + \r^{2}} + g^{0} \Big[ -2 \xi^{2} C_{\m\n} \del_{\n} K_{1} + 2 \bar\eta^{2} \r^{2} C_{\m\n} \del_{\n} K_{2} + 4 \r^{2} C_{\m\n} \del_{\n} K_{3} \Big] T^{4} \cr
& \l^{\a} = g^{-1} \Big[ {g \over 2} {{\s_{\m\n}}^{\a}}_{\b} \Big( \xi^{\b} -
\bar\eta_{\dot\gamma} {{\bar\s_{\r}^{\dot\gamma
\b}}}(x-x_{0})^{\r} \Big) {F_{\m\n}^{(0)}} \Big] \cr
& \bar\l_{\dot{\a}} = 0 \,.
}}

\subsec{Instanton solution}
The instanton solution is the same as
in the undeformed case. Starting now with ${F_{\m\n}^{(0),+}}=0$
and $\bar\l = \l = 0$, we note that in this case the Dirac
operator $\sigma^{\m}D_{\m}$ has non-trivial zeromodes for
$\bar\l$ whereas $\bar\sigma^{\m}D_{\m}$ has none, and thus
$\l=0$. Once again $\bar\l^{(1)}$ are given by the usual fermions
required by the index theorem. The difference with the previous
case is that now, as $\l$ has to be zero at this order in
Grassmann variables the equations of motion do no acquire a
fermion source term, and the initial bosonic solution remains a
solution. We can then write the full solution as
\eqn\fullinstanton{\eqalign{ & A_{\m} = g^{-1} { {2
{\bar\eta}_{\m\n}^{a}T^{a}(x-x_{0})_{\n}} \over (x-x_{0})^{2} +
\r^{2}} \cr & \l^{\a} = 0 \cr & \bar\l_{\dot{\a}} = g^{-1} \Big[
{g \over 2} {\bar\s}_{\m\n{\dot\a}{\dot\b}} \Big(
\bar\xi^{\dot\b} - \eta_{\gamma} {{\s_{\r}^{\gamma
\dot\b}}}(x-x_{0})^{\r} \Big) F_{\m\n}^{(0)}\Big] \,.}}


\newsec{The classical action}

In order to do a semi-classical calculation around the (anti-)instanton background, we would like to substitute the solutions found in the previous section into the classical action given in \ACT\ . We consider anti-instantons and instantons in turn.

\subsec{Anti-instanton} To find the classical action for the
anti-instanton solution we perform the
Bogomol'ny trick of \imA\ and write the action as \eqn\L{
S_{bos}= - \int d^{4}x {\rm Tr} \Big(F_{\m\n}^{-} - {i\over 2} g
C_{\m\n} \l \l \Big)^{2} - {1\over 4} \int {d^{4}x} {\rm Tr}
\e^{\m\n\rho\sigma} F_{\m\n} F_{\rho\sigma} \,.} The first term is
saturated for the soution found above, and thus the contribution
to the classical action comes entirely from the Chern-Simons-like
term. However one has to be aware of the fact that the field
strengths now depend on the Grassmann variables, and thus the
lagrangian will itself depend on fermionic coordinates.

To calculate the topological term we find it convenient to
rewrite it as \eqn\M{ -{1\over 4} \e^{\m\n\rho\sigma} F_{\m\n}
F_{\r\s} = - {1\over 2}F_{\m\n}(F_{\m\n} - 2F_{\m\n}^{-}) = -{1\over 2}
(F_{\m\n}F_{\m\n} - 2F_{\m\n}F_{\m\n}^{-}) \,.} We have to recall
that for $a$ in $SU(2)$ $F_{\m\n}^{a}$ is the usual
anti-instanton field strength such that $F_{\m\n}^{a,-}=0$ , and
$F_{\m\n}^{4,-} = {\xi}^2 {F_{\m\n}^{(\xi)^{2}4,-}} +
{\bar\eta^2} F_{\m\n}^{(\bar\eta^{2})4,-} + F_{\m\n}^{(\bar\eta \xi)4,-}$. The total anti-instanton action
will have contributions \eqn\N{\hskip-5cm
 -{1\over 4}{\rm Tr}
\e^{\m\n\r\s}F_{\m\n}  F_{\r\s} = - {1 \over 2} {\rm Tr}
(F_{\m\n}F_{\m\n} - 2F_{\m\n}F_{\m\n}^{-}) = }
$$\hskip2cm
= {1\over 4} \Big[F_{\m\n}^{a}F_{\m\n}^{a} + \left(F_{\m\n}^{(\bar\eta \xi),4} F_{\m\n}^{(\bar\eta
\xi),4} - 2 F_{\m\n}^{(\bar\eta \xi),4} F_{\m\n}^{(\bar\eta
\xi),4-}\right) +
$$\hskip2cm
$$
+ (\bar\eta^{2} \xi^{2})\left(2
F_{\m\n}^{(\bar\eta^2),4}F_{\m\n}^{(\xi^2),4}
-2\left(F^{(\bar\eta^2),4-}_{\m\n}F_{\m\n}^{(\xi^2),4}+F^{(\bar\eta^2),4}_{\m\n}F_{\m\n}^{(\xi^2),4-}\right)\right)\Big]\,.
$$
From the expressions for $K_1$, $K_2$, $K_3$ we can calculate this
quantity. Recalling \eqn\O{ \xi^2 F_{\m\n}^{(\xi^2),4} = 2
\xi^2 (C_{\n\r} \del_{\m} \del_{\r} - C_{\m\s} \del_{\n}
\del_{\s})K_{1} }
$$
\bar\eta^2 F_{\m\n}^{(\bar\eta^2),4} = -2 \r^{2} \bar\eta^2 (C_{\n\r} \del_{\m} \del_{\r} - C_{\m\s} \del_{\n} \del_{\s})K_{2}
$$
$$
F_{\m\n}^{(\bar\eta \xi),4} = -4 \r^{2} (C_{\n\r} \del_{\m} \del_{\r} - C_{\m\s} \del_{\n} \del_{\s})K_{3}
$$
we find \eqn\Q{ {1\over 2}\e^{\m\n\r\s}F_{\m\n}^{(4)}F_{\r\s}^{(4)}={3072\r^6 (2x^4-3\r^2 x^2)\over(x^2 +\r^2)^6}|C|^2 \bar\eta^2\xi^2\,.}  Putting
everything together the classical action for the anti-instanton is
\eqn\caain{
S_{anti-instanton}=\int d^4 x \left[ {96\over g^2}{\r^4\over(x^2+\r^2)^4} +{3072\r^6 (2x^4-3\r^2 x^2)\over(x^2 +\r^2)^6}|C|^2 \bar\eta^2\xi^2\right] = {8\pi^2\over g^2}\,.}
We then see that, although the $U(1)$ anti-intanton charge density is non vanishing, only the $SU(2)$ part contributes to the total topological charge. This is in accordance with the index theorem.
\subsec{Instanton}

Because in the instanton solution to the equations of motion one has $\l=0$ the classical instanton action does not suffer any modification from the usual $N=1$ super Yang-Mills. We then have
\eqn\instclas{
S_{instanton}= {8 \pi^{2} \over g^{2}}\,.
}

\newsec{Path integral measure and semi-classical approximation}

Before we proceed to set up a semi-classical calculation around the (anti-) instanton solutions found in previous sections, we review the conventional approach to better contrast the new elements which arise in our situation.

The general procedure consists in splitting the fields in the path integral into a classical part which satisfies the classical equations of motion and a quantum part which describes the fluctuations around the classical solution. After fixing a gauge, one plugs the field expansion into the gauge-fixed action and keeps terms up to second order in quantum fluctuations. This yields a product of determinants corresponding to the different fields in the path integral. A characteristic feature of instanton calculations is that the quadratic operators corresponding to these determinants have zero modes which must be treated separately in order for the path integral to give a sensible result. The presence of these zero modes is due to a degeneracy of lowest-energy configurations, which is parameterized by a set of collective coordinates. One must choose a gauge to fix this degeneracy and trade in the integration over zero-modes in the path integral for an integration over the collective coordinates. In the process one picks up a jacobian factor which determines the measure of integration over the collective coordinates.

To see this in a little more detail consider the generic field expansion
\eqn\expansion{
\phi_{m}^{M}(x)= \phi_{m}^{M}(x;X) + \delta\phi_{m}^{M}(x;X)
}
where $\phi(x;X)$ is a classical background with degeneracy parameterized by collective coordinates denoted generically by $X^{A}$ (these encompass both fermionic and bosonic collective coordinates), and $\delta\phi(x;X)$ is the quantum fluctuation. The zero modes are then $\delta_{A}\phi_{n}(x)$, where $\delta_{A}$ denotes differentiation with respect to $X^{A}$. The action is then expanded to quadratic order in the fluctuations, $S=\delta\phi_{n}^{N} {\cal O}_{nm}^{NM}\delta\phi_{m}^{M}$. Expanding the fluctuations in terms of a complete set of eigenvectors of ${\cal O}$
\eqn\qfl{
\delta\phi_{n}^{M} = \sum_{A} \xi^{A} \delta_{A} \phi_{n}^{M} + \tilde{\phi}_{n}^{M}
}
where $\delta_{A} \phi_{n}^{M}$ are the zero-modes and $\tilde{\phi}_{n}^{M}$ are the non-zero eigenvectors, the path integral measure becomes
\eqn\pmeas{
\int [d\phi_{m}^{M}] = \int \Big[ \sqrt{{\rm Sdet} \, g_{AB}(X)} \prod_{A} {d\xi^{A}\over {\sqrt {2\pi}}} \Big] [d\tilde \phi_{m}^{M}]}
where $g_{AB}= k \int d^{4}x {\rm} Tr \delta_{A}\phi_{r}^{M} \delta_{B}\phi_{r}^{M}$ is the suitably normalized metric of inner products of the zero-modes.

The next step is to fix the gauge for both non-zero and zero-modes, and trade $d\xi^A$ for $dX^a$ in the integration over the zero-modes. To do that one performs a BRST quantization inserting unity into the path integral in the form \foot{We are being schematic here. There are subtleties concerning the aymptotics of the gauge transformations, which entangle the gauge fixing for zero and non-zero modes. For more details see \lref\AmatiWU{
D.~Amati and A.~Rouet,
``Renormalization Of Yang-Mills Theory Developed Around An Instanton,''
Nuovo Cim.\ A {\bf 50}, 265 (1979).
} \AmatiWU .}
\eqn\faddeevpopov{
1=\int \prod_{A}[dX^{A}][d\Omega(x)] \delta(G(\tilde\phi^{\Omega})) \prod_{n} \delta(f_{A}(X)) {\rm det} \Big| {\delta G(\tilde\phi^{\Omega})\over \delta \Omega} \Big| {\rm Sdet} \Big| {\delta f_{A}\over \delta X^{B}} \Big|}
where $G(\tilde\phi)=D_{m}\tilde\phi_{m}$ (background gauge) and $f_{A}= k \int d^{4}x {\rm} Tr \delta\phi_{m}^{M} \delta_{A}\phi_{m}^{M}= \sum_{A} \xi^{B}g_{AB}$.  $\delta(G)$ fixes the gauge for the non-zero modes, and ${\rm det} \Big| {\delta G(\phi^{\tilde\Omega})\over \delta \Omega} \Big|$ gives the usual Faddeev-Popov determinant.  For the zero-modes $\delta(f_{A}(X))$ enforces $\xi^{A}=0$, as $g_{AB}$ is invertible. Moreover to leading order in g, ${\delta f_{A}\over \delta X^{B}}=g_{AB}$ so that
\eqn\fpdos{
{\rm Sdet} \Big|{\delta f_{A}\over \delta X^{B}} \Big| = {\rm Sdet  } [g_{AB}]
}
The measure for the gauge-fixed action becomes
\eqn\gfam{
\int \Big[ \sqrt{{\rm Sdet} \, g_{AB}(X)} \prod_{A} {dX^{A}\over {\sqrt {2\pi}}} \Big] [d\tilde \phi_{m}^{M}]{\rm det}\, (-D^{(0)})^2}
Note that one can introduce additional anticommuting ghosts $\bar{c}^{(0)a}$ and $c^{(0)b}$ for the bosonic zero-modes, and commuting ghosts $\bar{\gamma}^{(0)\a}$ and $\gamma^{(0)\b}$ for the fermionic zero-modes, and write
\eqn\fan{
\sqrt{{\rm Sdet}\, g_{AB}} = \int [d \bar{c}^{(0)}][dc^{(0)}][d \bar{\gamma}^{(0)}][d \gamma^{(0)}] {\rm exp}[{1\over 2}\bar{c}^{(0)a}g_{ab} c^{(0)b}]{\rm exp}[{1\over 2}\bar{\gamma}^{(0)\a}g_{\a\b} \gamma^{(0)\b}]}
assuming orthogonality between bosonic and fermionic zero-modes.
We now see how these considerations apply to our situation.

\subsec{Anti-instanton}
The splitting into background plus fluctuation of the fields in the path integral takes the form
\eqn\bkgr{\eqalign{
& A_{\m}^{a} = {1\over g} A_{\m}^{(0),a} + Q_{\m}^{a},~~~~~~~~ A_{\m}^{4} = A_{\m}^{(1),4} + Q_{\m}^{4} \cr
&\bar\l = \bar{q},~~~~~~~~~~~~\l = g^{-1} \l^{(0)}+q
}}
with the expressions for the different fields given in \fullantiinstanton. We now must compute the quadratic part in the fluctuations of the classical lagrangian. However, we should stress that the background \bkgr\ is non-conventional, insofar as it includes fermions. Usually fermion zero-modes are not included in the classical background, and are treated in perturbation theory \foot{We acknowledge P. van Nieuwenhuizen and S. Vandoren for communicating to us their preliminary results \instquatt\ for $N=4$ SYM.}. 
The main difference with the conventional treatment boils down to the fact that the background fermions induce additional fermion-boson and fermion-fermion couplings in the quadratic expansion of the classical action. In the one-loop calculation one is then forced to compute the {\it superdeterminant} of these fields.

To perform the computation we choose a $U(2)$ background gauge fixing
$$
D_{\m}^{(0)}Q_{\m} = 0
$$
which adds \eqn\add{ {\cal L}_{g.f.} + {\cal L}_{ghost} = {\rm
Tr} \Big[ (D^{(0)}_{\m}Q^{a}_{\m})^{2} -2bD^{(0)2}c\Big]} to the
action. The gauge-fixed action action, up to quadratic
fluctuations, can be written as \eqn\squadA{ S_{quad}=\int d^{4}x
\Big[{\cal L}_{1} + {\cal L}_{g.f.} + {\cal L}_{ghost}\Big] }
where \eqn\squadB{ {\cal L}_{1} + {\cal L}_{g.f.}  = \left(
\matrix{Q_{\n} & q^{\b} & \bar{q}_{\dot{\b}}} \right)  \left(
\matrix{{\cal A}_{\n \m} & {{\cal B}_{\n}}^{\a} & {\cal C}_{\n
\dot{\a}} \cr {\cal D}_{\b \m} & {{\cal E}_{\b}}^{\a} & {\cal
F}_{\b \dot{\a}} \cr {{\cal G}^{\dot{\b}}}_{\m} & {\cal
H}^{\dot{\b}\a} & {{\cal I}^{\dot{\b}}}_{\dot{\a}}} \right)
\left( \matrix{Q_{\m} \cr q_{\a} \cr {\bar{q}}^{\dot{\a}}}
\right)} and we have suppressed gauge indices for clarity. The
different elements of this matrix can be read from \eqn\quanonc{
{1\over 2} {\rm Tr} F_{\m\n}F_{\m\n} \Big|_{quad} =  \Big[ -{1
\over 2} (D_{\m}^{(0)}Q_{\n}^{a})^{2} + {1 \over 2} (D_{\m}^{(0)}
Q_{\m}^{a})^{2} - g \epsilon_{abc} Q_{\n}^{a} F_{\m\n}^{(0),b}
Q_{\m}^{c} \Big] + }
$$\hskip-1.8cm
+ \Big[ - {1 \over 2} (\del_{\m} Q^{4}_{\n})^{2} + {1 \over 2}(\del_{\m} Q^{4}_{\m})^{2} \Big]
$$
$$\hskip -2cm
2 {\rm tr} \l \not\!\!D \bar\l |_{quad}= -\bar{q}^{4}
\bar\s^{\m}{\del}_{\m}q^{4} - q^4 \s^\m \del_\m \bar{q}^4  -
\bar{q}^{a}(\bar\s^{\m}D_{\m}^{(0)}q)^{a} - q^a (\s^\m
D_\m^{(0)}\bar q)^a$$
$$
-  \left( \bar{q}^{a} \bar{\s}^{\m}[Q_{\m},\l^{(0)}]^{a}+\l^{(0),a} \s^{\m} [Q_\m , \bar q]^a\right)
$$
and \eqn\quadratic{\hskip-3.2cm i g {\rm Tr} \Big[ C^{\m\n}
F_{\m\n} \l \l \Big] \Big|_{quad}  = {1\over 4} \Big[  C_{\m\n}
\del_{[\m} A^{(1),4}_{\n]} (q^{4} q^{4} + q^{a} q^{a})\Big] + }
$$\hskip+2.5cm
+ {1\over 2} \Big[C_{\m\n} \left( \Big( \del_{[\m}
Q_{\n]}^{a}+ [A_\m^{(0)},Q_\n ]^{a} \Big)q^{4} \l^{(0),a} +
\del_{[\m}Q^{4}_{\n]} q^{a} \l^{(0),a}\right) \Big]
$$
\eqn\q{ g^{2} {\rm Tr} \Big[ {|C|^2 \over 4} (\l \l)^{2} \Big]
\Big|_{quad} = {|C|^2 \over 4} \Big[ {1\over 4} (\l^{(0),a}
\l^{(0),a} )(q^{b}q^{b}) +{1\over 2}(\l^{(0),a}q^{a} \l^{(0),b
}q^{b} )\Big]}
$$
$$
 The expansion of the superdeterminant
\squadB\ can be done systematically using \eqn\sd{ {\rm Sdet}
\Big( \matrix{ X_{bb} & Y_{bf} \cr W_{fb} & Z_{ff}} \Big)
 = {{\rm det X_{bb}} \over {\rm det}(Z_{ff}-W_{fb}X_{bb}^{-1}Y_{bf})}}
In our case $X_{bb}={\cal A}_{\n\m}$ is the usual bosonic quadratic operator
 for $U(2)$ (super) Yang-Mills ,
 $Z_{ff}= \left( \matrix{{{\cal E}_{\b}}^{\a} & {\cal F}_{\b \dot{\a}} \cr {\cal H}^{\dot{\b}\a}
  & {{\cal I}^{\dot{\b}}}_{\dot{\a}}} \right) = \Delta_{D}+\Delta_{E}$.  $\Delta_{D}$
  is the usual operator for adjoint Dirac fermions in $U(2)$  (super) Yang-Mills
  and $\Delta_{E}$ encodes the additional fermion-fermion couplings arising
   from the first line in \quadratic\  and from \q . We see that ${{\cal I}^{\dot{\b}}}_{\dot{\a}}=0$. Finally $Y_{bf}=\left(\matrix{{{\cal B}_{\n}}^{\a} & {\cal C}_{\n \dot{\a}}} \right)$ and $W$ its fermionic transpose. We can expand the logarithm of the superdeterminant as
\eqn\logsup{\hskip-3.8cm
{\rm log}\,{\rm Sdet} \left( \matrix{ X_{bb} & Y_{bf} \cr W_{fb} & Z_{ff}} \right) = {\rm log}\,{\rm det}{\cal A}_{\m\n} - {\rm log}\,{\rm det}\Delta_{D} -
}
$$\hskip1.5cm
 - {\rm Tr}\,{\rm log} \left(1+\Delta_{D}^{-1}\Delta_{E}-\Delta_{D}^{-1} \left( \matrix{{\cal D}_{\b \m} \cr {{\cal G}^{\dot{\b}}}_{\m}} \right) ({\cal A}_{\m\n})^{-1} \left( \matrix{{{\cal B}_{\n}}^{\a} & {\cal C}_{\n \dot{\a}}} \right) \right)
$$
The first two terms in \logsup\ give the same contributions as in $N=1$  $SU(2)$
super Yang-Mills, as the $U(1)$ part in these terms yields only an infinite
constant which is canceled  once we normalize with respect to the vacuum.
 The calculation is standard, but we reproduce it here for completeness.
  Integration over $A_{\m}^{a}$  yields
\eqn\detbosons{
[{\rm det}' \, {\cal A}_{\m\n}]^{-1/2}
}
where ${\cal A}_{\m\n} = - (D^{(0)})^{2} \delta_{\m\n} - 2 F^{(0)}_{\m\n}$ , and the prime indicates that the determinant has to be amputated because, as usual, it has zero-modes.
Integration over the $SU(2)$ fermions gives
\eqn\detfermions{
{\rm det}\,{\Delta_{D}}={\rm det}\left(\matrix{0 & \sigma^{\m}D_{\m} \cr \bar{\sigma}^{\m}D_{\m} & 0}\right)=[{\rm det}' \, \Delta_{-}]^{1/4} [{\rm det}\, \Delta_{+}]^{1/4}
}
Here $\Delta_{+} = - \bar\sigma^{\m} D_{\m} \sigma^{\n}D_{\n} = -D^{2}$ is the hermitian operator for the $SU(2)$ $\bar\l$ fluctuations (with the same spectrum as $\sigma^{\m}D_{\m}$). As there are no $\bar\l$ zeromodes the determinant is the full one. On the other hand $\Delta_{-} = - \sigma^{\m} D_{\m} \bar\sigma^{\n}D_{\n} = -D^{2} - \sigma_{\m\n}F_{\m\n}^{(0)}$ has zeromodes, and the determinant has to be amputated. The spectrum of non-zero eigenvalues of $\Delta_{+}$ and $\Delta_{-}$  is the same.
For the ghosts one gets ${\rm det} \Delta_{ghosts}$, where $\Delta_{ghosts} = - D^{2}$.

It is a standard fact that
\eqn\cancellation{\eqalign{
& {\rm det'} {\cal A}_{\m\n} = [{\rm det'} \Delta_{-}] ^{2}\cr
& {\rm det} \Delta_{ghosts} =[ {\rm det} \Delta_{+} ]^{1/2}
}}
Using this the total product of determinants is given by
\eqn\totdet{
\left[{{\rm det}\Delta_{+}\over {\rm det'}\Delta_{-}}\right]^{3/4}
}
which is formally one, since as mentioned before both determinants have the same spectrum of non-zero eigenvalues. One has to introduce a regularization scheme to make sense of these determinants, however. The usual procedure is to use a Pauli-Villars regulator mass. With this the total product of determinants picks up a factor of $\m^{n_{b}-{1\over 2}n_{f}}=\m^{6}$ where $n_{b(f)}$ is the number of bosonic (fermionic) zero-modes.

The new part of the calculation comes from the second line in
\logsup. One has to expand the logarithm in powers of the
background fields and keep gauge invariant combinations. The
possible one-loop diagrams contributing have been analyzed in
\AlishahihaKG \foot{In the computation that follows one can use conventional Background Field formalism techniques, with the exception that the propagator $\Delta_D^{-1}$ contains no zero-modes. However, we will only need the UV divergent part of the Feyman diagrams, and therefore the amputated propagators can be well approximated by the usual background field ones in this regime.}. The non vanishing one-loop Feynman diagrams come
in two different topologies (see figs. 3 and 6 in \AlishahihaKG).
We now show how these diagrams appear from the super-determinant
expansion. The first diagram contributes to the renormalization of
the $A\l\l$ vertex and contains a loop made of two fermionic and
one bosonic propagator, in which the bosonic propagator runs
between two external $\l$, and the fermionic between external $A$
and $\l$. The modified C dependent vertex can be any vertex of
the diagram. Using the notation of \AlishahihaKG\  the diagrams
$(1_c,2,3),(1,2_c,3),(1,2,3_c)$ all contribute (the index "c"
specifies the position of the modified vertex). As usual in
background field formalism only a reduced set of diagrams
contribute. For this topology it is easy to see, using the
background Feynman rules, that only the diagram $(1,2,3_c)$
contribute with a background $U(1)$ photon entering vertex 3, 2
$SU(2)$ fluctuation fermions and one $SU(2)$ fluctuation photon
circulating in the loop. Putting $x=\Delta_{D}^{-1}\Delta_{E}$,
$y=-\Delta_{D}^{-1} \left( \matrix{{\cal D}_{\b \m} \cr {{\cal
G}^{\dot{\b}}}_{\m}} \right) ({\cal A}_{\m\n})^{-1} \left(
\matrix{{{\cal B}_{\n}}^{\a} & {\cal C}_{\n \dot{\a}}} \right)$
and expanding the logarithm  we see that in our approach\foot{We
are using a simple non-supersymmetric background gauge in which,
unlike \AlishahihaKG, there are no couplings between ghosts and
fermions.}  the three vertex diagram arises from the term $xy$.
Indeed in $xy$ one finds the term \eqn\newpart{
{(\Delta_{D}^{-1})^{\dot\a}}_{\b}{(\Delta_{E})^{\b}}_{\gamma}{(\Delta_{D}^{-1})^{\gamma}}_{\dot\delta}{{\cal
G}^{\dot\delta}}_{\m}{\cal A}^{-1}_{\m\n}{\cal C}_{\n\dot\a}}
with two background fermions $\l^a\l^a$ and one background
$A_{\m}^{4}$ (note that terms with an $A_{\m}^{a}$ background
vanish due to the anti-selduality of $C$). All the other terms in
$xy$ do not give diagrams consistent with the Feynman rules and
therefore do not contribute.  We can also draw a new diagram which
is identical to $(1,2,3_c)$ apart from substituting the modified
vertex with a new vertex originating from the term $-|C|^2/16
(q^aq^a)(\l^{(0)b}\l^{(0),b})$ present in
$-|C|^2/4(\l\l)^2_{quad.}$. The new diagram cancels exactly the
previous one. Indeed the modified vertex in $(1,2,3_c)$ comes
from $-1/4
C_{\m\n}\partial_{[\m}A_{\n]}^4q^aq^a=1/16|C|^2\l^{(0),a}\l^{(0),a}q^bq^b
$ where we used the deformed anti-instanton equation
$F_{\m\n}^{(-),4}=-1/4C_{\m\n}\l^{(0),a}\l^{(0),a}$.

By power counting at the vertices (and accounting for the
$g$-scalings of the different background fields) this diagram
goes as $|C|^2$. Moreover, the external field structure
implies that it is proportional to $\bar\eta^2 \xi^2$. The other
non-vanishing diagram is the one that renormalizes the $\l\l\l\l$
coupling and has four external fermions, a loop with two
fermionic propagators and two bosonic propagators. Using the
background Feynman rules we find that, in agreement with
\AlishahihaKG, the only consistent diagrams are $(1_c,2_c,3,4)$
and $(1,2,3_c,4_c)$ with one U(1) photon connecting the
C-modified vertices and SU(2) fermions circulating in the loop.
This diagrams come from the $y^2$ term of the super-determinant
expansion. For instance the diagram $(1_c,2_c,3,4)$ arises from
\eqn\newpartB{ {(\Delta_{D}^{-1})^{\a}}_{\dot\b} {{\cal
G}^{\dot\b}}_{\m}({\cal A}^{-1})_{\m\n}{\cal
C}_{\n\dot\gamma}{(\Delta_{D}^{-1})^{\dot\gamma}}_{\delta}{{\cal
D}^{\delta}}_{\r}({\cal A}^{-1})_{\r\s}{{\cal B}^{\s}}_{\a}.} The
contribution of this diagram is also zero because we have four
fermions in the background and $(\l^{(0),a}\l^{(0),a})^2=0$.
Once again this goes as $|C|^2 \bar\eta^2 \xi^2$. It is also
possible to check that all the other terms coming from the
super-determinant expansion do not generate consistent diagrams.
Therefore the one-loop effective action is zero and the zero
modes remain unlifted to this order in perturbation
theory. Much as in the $N=1$ case, the supersymmetries left unbroken by both the anti-instanton and the C-deformation are still effective in compensating the bosonic and fermionic fluctuations in the one-loop (super)determinant.

Finally, the last part of the calculation corresponds to the integration the modes that were amputated in the determinants of the first line in \logsup, namely the zero-modes of $A_{\m\n}$ and $\Delta_{-}$. The calculation reduces to the computation of the superdeterminant \fan\ for ordinary $N=1$ SYM.
\eqn\measureA{
\sqrt{{\rm Sdet}\,g_{AB}}|_{\rm N=1 SYM} = \Big[ \int db_{a}db_{b} e^{- b_{a} <{\del A_{\m} \over \del b_{a}}| {\del A_{\m} \over \del b_{b}}> b_{b}} \Big]^{-1/2} \Big[ \int df_{\a}df_{\b} e^{- f_{\a} <{\del \l^{\gamma} \over \del f_{\a}}| {\del \l_{\gamma} \over \del f_{\b}}> f_{\b}}\Big]^{-1/2}
}
$$\hskip-3cm
= 2^{10} \pi^{6} g^{-8} \r^{3} \Big({g^{2}\over 32\pi^{2}} \Big) \Big({g^{2}\over 64\pi^{2}\r^{2}} \Big)
$$
In the formula above $b_{a}$ stands for bosonic coordinates, namely $x_{0},\r$ and gauge orientations, and $f_{\alpha}$ stands for $\bar\eta, \xi$. The normalization for the inner products is
\eqn\innpr{\eqalign{
& \left<{\del A_{\m} \over \del b_{a}}\Big| {\del A_{\m} \over \del b_{b}}\right>= g_{ab}=-{2\over g^{2}} \int d^{4}x {\rm Tr}\, \delta_{a}A^{(0)}_{\m} \delta_{b}A^{(0)\m}\cr
&  \left<{\del \l^{\gamma} \over \del f_{\a}}\Big| {\del \l_{\gamma} \over \del f_{\b}}\right> =g_{\a\b}=-{2\over g^2} \int d^{4}x {\rm Tr}\, \delta_{\a}\l^{(0)}_{\gamma} \delta_{\b}\l^{(0)\gamma}
}}
Boson and fermion zero-modes are orthogonal. Putting everything together the total semi-classical partition function for the anti-instanton is
\eqn\measureB{
{\cal Z} =
2^{10} \pi^{6} g^{-8} \m^{6}{\rm exp} \Big[ - \Big( {8\pi^{2}\over g^{2}}  - i{\theta} \Big) \Big]
}
$$
\int d^{4}x_{0} \int \r^{3} d\r \int d^{2}\xi \Big({g^{2}\over 32 \pi^{2}} \Big) \int d^{2} \bar\eta \Big({g^{2}\over 64 \pi^{2} \r^{2}} \Big)
$$

As a final remark we notice that the gluino condensate is
unchanged with respect to the familiar N=1 result. This is due to
the fact that volume form in the anti-instanton moduli space does
not get C-corrections.
\subsec{Instanton}
In the case of the instanton the g-expansion of the different fields is as follows
\eqn\antiins{\eqalign{
& A_{\m}^{a} = g^{-1} A_{\m}^{(0),a} + Q_{\m}^{a},~~~~~~A_{\m}^{4} = Q_{\m}^4 \cr
& \bar\l = g^{-1} {\bar\l}^{(0)} + \bar{q},~~~~~~~\l = q \cr
}}
The expansion, up to quadratic terms in fluctuations, of the C-deformed part of the action is then
\eqn\quadact{\eqalign{
& ig {\rm Tr} C_{\m\n}F_{\m\n} \l \l |_{quad} = {1\over 2} C_{\m\n}F_{\m\n}^{(0),a} q^{4} q^{a} \cr
& - g^{2} {\rm Tr} {|C|^{2}\over 4} (\l \l)^{2} |_{quad} = 0
}}
The undeformed part of the action also acquires a new coupling with respect to usual $U(2)$ $N=1$ SYM when we substitute the background fermion $\bar\l^{(0)}$. It is given by
\eqn\elb{
g^{1/2}\left(\bar\l^{(0),a} \bar\sigma^\m [Q_\m , q]^a+q^{a}\sigma^\m [Q_\m, \bar\l^{(0)}]^a \right)}
however, in this case it is not strictly necessary to incorporate the $\bar\l$ fermions into the classical background, as the classical instanton action does not depend on the background fermions. We can then follow the usual prescription and treat them in perturbation theory. We should however take into account the extra bilinear fermion coupling in \quadact. Writing the fermions in Dirac form it can be seen to contribute a (gauge off-diagonal) mass term, which can be treated as a perturbation to the usual Dirac operator $\Delta_{D}$. However, because of the chirality of the two fermions of this new vertex, it does not contribute to the determinant. The analysis can be reduced to standard techniques and we will not further pursue it here.


\newsec{Outlook}
The results in this paper should be seen as a preliminary step to
an instanton calculation with matter included. Indeed, for the
weak coupling semi-classical approximation to be self-consistent,
 in the moduli space integration one should introduce an infrared
  cutoff to prevent the coupling constant from becoming strong.
  The one-loop behavior of $g$ for $N=1/2$ supersymmetric theories
  is the same as in $N=1$ SYM, so one inevitably runs into strongly
  coupled regions. It is hoped that the Higgs field $VEV$ will provide
   the infrared cutoff, just as in the ordinary case. 
       (It would be interesting
       to explore higher-loop renormalizations, using explicit supergraph techniques
       for example, as well as genuinely non-perturbative consequences our results
        could have for $N=1/2$ supersymmetric theories.)

A different direction would be to explore whether more general
(anti-) instanton solutions to the C-deformed equations of motion exist, and what relevance these could have both at the mathematical and the physical level. The case of multi-instantons immediately comes to mind.

Another very interesting application of the present analysis would be the extension to N=2 SYM
(see for example \lref\Fer{S.~Ferrara and E.~Sokatchev,
{\it Non-anticommutative N = 2 super-Yang-Mills theory with singlet deformation,} arXiv:hep-th/0308021;
E.~Ivanov, O.~Lechtenfeld and B.~Zupnik,
{\it Nilpotent deformations of N = 2 superspace,}
arXiv:hep-th/0308012.

} \Fer)  where the complete many instanton computation can be
performed along the lines of
\lref\Nekra{
N.~A.~Nekrasov,
{\it Seiberg-Witten prepotential from instanton counting,}
arXiv:hep-th/0206161.
} \Nekra\ using the methods of equivariant cohomology.

Finally, a dual picture can be constructed. Using the analysis of Ooguri and Vafa one can see that the
supersymmetry can be restored by changing the commutation properties of the gluino $\bar\l$
\eqn\ooA{
\{ \bar\l^{a}_{\a}, \bar\l^{b}_{\b} \} = C_{\a\b} \delta^{ab} \,.
}
However, this prescription has to be handled at the quantum level in the
process of quantizing the gluinos. This implies that the system is constrained and
therefore has be carefully discussed. We notice however that there is a similarity in the
eqs. of Imaapur \imA\ and the constraints on the gluinos. We see that the eqs. \ooA\ coincides
with \imA\ if $F^{a,+}_{\mu\nu} \propto C_{\mu\nu}$.

\newsec{Acknowledgements}

We thank Martin Ro\v cek and Peter van Nieuwenhuizen for useful discussions. We also wish to express our gratitude to the organizers of the Simons Workshop in Mathematics and Theoretical Physics at Stony Brook for providing a stimulating atmosphere which encouraged this project. This work is partially supported by NSF grant PHY-0098527.

\vskip .5cm

{\bf Addendum}
After completion of the first version of this paper, but before submission to the archives,
\lref\imI{A. Imaanpur, {\it Comments on Gluino Condensates in N=1/2 SYM Theory}, hep-th/0311137} \imI\
appeared, there is a partial overlap with our results. Also, in the first version, it was erroneously claimed that the $U(1)$ topological charge was not vanishing, as implicitly pointed out in 
\lref\BrittoUV{
R.~Britto, B.~Feng, O.~Lunin and S.~J.~Rey,
{\it U(N) instantons on N = 1/2 superspace: Exact solution 
and geometry of moduli space,}
arXiv:hep-th/0311275.
} \BrittoUV \foot{We acknowledge them for making us realize it.}. We believe, however, that the C independence of gluino condensates, and of the volume of the deformed $U(2)$ anti-instanton moduli space, pointed out in that previous version, are independent of that flaw. Moreover, the general iterative method of solving the equations of motion in fermionic quasi-collective coordinates, and the dependence of the topological charge density on these quasi-collective coordinates and C, are of course unaffected. 


\listrefs

\bye